\documentclass[10pt,a4paper]{article}
\usepackage[utf8]{inputenc}
\usepackage{amsmath}
\usepackage{amsfonts}
\usepackage{amssymb}
\usepackage{amsthm}

\usepackage{graphicx}
\usepackage{jheppub}

\title{Semiclassical expansion of the Bethe scalar products in the XXZ spin chain}
\author{Constantin Babenko}

\affiliation{Laboratoire de Physique Théorique et Hautes Énergies, Université Pierre et Marie Curie\\
4 Place Jussieu, 75004 Paris, France}

\emailAdd{cbabenko@lpthe.jussieu.fr}

\abstract{In this note we adress the problem of performing the semiclassical expansion of the scalar product of Bethe states in the case of the XXZ spin chain. Our approach closely follows the one developped in \cite{Kostov2}: after expressing the scalar products in terms of a central quantity - the $B$ functional, we reinterpret it as a grand partition function of a gas of particles with pairwise interaction potential. This work has been done as part of the author's master thesis under the supervision of Ivan Kostov and Didina Serban at \textit{Institut de Physique Théorique} (CEA Saclay) in September 2015 - February 2016, and present a consice summary of the results obtained at that time.}


\begin{document}
\maketitle

\begin{small}
\section{Position of the problem}
In this note we are concerned with some aspects of the scalar products of the XXZ spin chain. Spin chains provide one of the most famous example of integrable systems, namely systems which can be exactly soved and for which  the quantities of physical interest can be computed analytically. In the case of spin chains, the procedure that extracts the physical information bears the name of the Algebraic Bethe Ansatz and yields the so called Bethe vectors, which are the eigenvectors of the hamiltonian of the spin chain. Here we propose to study the scalar products between such vectors, in a special limit. 
In addition to be of intrisic interest for the study of integrable systems, this problem (and more broadly the understanding of the behaviour of spin chains) is also motivated by its applications to  the study of the AdS/CFT correspondence, where it has been shown since a decade that integrability plays a major role. 

First we define what we mean by semiclassical expansion of Bethe states. The system under study is the XXZ spin chain of lenght $L$ whoose $R$ matrix is given by :
\begin{equation}
R(u)
=
\begin{pmatrix}
\mathrm{sh}(u +i \eta) & 0 & 0 & 0\\
0 & \mathrm{sh}(u ) & \mathrm{sh}(i \eta) & 0\\
    0   &  \mathrm{sh}(i \eta)  & \mathrm{sh}(u ) & \\
    0   &   0  & 0 & \mathrm{sh}(u+i \eta ) 
\end{pmatrix}
\end{equation}
From the $R$ matrix we define the transfer matrix $T$ written in compact form as :
\begin{equation}
T(u)
=
\begin{pmatrix}
   A(u) & B(u) \\
   C(u) & D(u) 
\end{pmatrix}
\end{equation}
The Bethe vectors are then defined as multiple actions of the operators $B$ and $C$ on the vacuum of the system $|\Omega\rangle$ (bold leters will denote ordered sets of complex numbers : $\mathbf{a}=(a_{1},...,a_{N})$) :
\begin{equation}
| \mathbf{u} \rangle
:=
B(u_{1})...B(u_{M})|\Omega\rangle  \;\;\;\;\;\;
\langle \mathbf{v} |
:=
\langle\Omega| \; C(v_{1})...C(v_{M})
\end{equation}
where we assumed that the rapidity sets  $\mathbf{u}$ and $\mathbf{v}$ satisfy the Bethe equations ($\theta_{m}$ represents the inhomogenety located at the site $m$ of the chain) :

\begin{equation}
\prod_{k=1}^{M}
\frac{\mathrm{sh}( u_{j}-u_{k}+i\eta)}{\mathrm{sh}(u_{j}-u_{k}-i\eta)} 
=
-
\prod_{m=1}^{L}
\frac{\mathrm{sh}( u_{j}-\theta_{m}+i\eta)}{\mathrm{sh}(u_{j}-\theta_{m})} 
\end{equation}
The scalar product of Bethe vectors is then given by :
\begin{equation}
\label{innerprod}
\langle\mathbf{v}|\mathbf{u}\rangle
=
\langle\Omega|\prod_{k=1}^{M}C(v_{k})
\prod_{k=1}^{M}B(u_{k})|\Omega\rangle
\end{equation}
These scalar products will be our main object of  study. We aim to find the asymptotical expansion of  \eqref{innerprod}  in the semiclassical limit, ie in the limit $M\rightarrow\infty $, and our reasoning will closely follow the papers \cite{Kostov2} and \cite{Kostov1}, where the same problem has been resolved in the case of a XXX spin chain.
\\
The main results of this work are given by \eqref{final2}, which expresses  \eqref{innerprod} in terms of a specific quantity $B$, which should admit a particularly simple semiclassical expansion and \eqref{final3}/\eqref{final}, which are the first steps towards this expansion. We first start by proving the formula \eqref{final2}.

\section{The B functional}
We now consider two Bethe vectors : $|\textbf{v}\rangle $ and  $|\textbf{u}\rangle $.  As it is explained in \cite{Kostov1},  it is possible to bring the scalar product  to the form :
\begin{equation}
\label{innerprodxxz}
\langle \textbf{v}|\textbf{u}\rangle
=
\prod_{j=1}^{M}
\mathcal{A}(v_{j})\mathcal{D}(u_{j})
S_{\textbf{u},\textbf{v}}
\end{equation}
where the functions
\begin{align}
&\mathcal{A}(u)=\prod_{m=1}^{L}\mathrm{sh}(u-\theta_{m}+i\eta)  \;\;\;\;\;\;
\mathcal{D}(u)=\prod_{m=1}^{L}\mathrm{sh}(u-\theta_{m})
\end{align}
correspond to the vacuum eigenvalues of $A(u)$ and $C(u)$, and  the quantity $S_{\textbf{u},\textbf{v}}$ is defined by :
\begin{align}
S_{\textbf{u},\textbf{v}}
&=
\frac{\mathrm{det}_{jk}\Omega(u_{j},v_{k})}
{\mathrm{det}_{jk}\frac{1}{\mathrm{sh}(u_{j}-v_{k}+i)}} \;\;\;\;\;\;\;\;
\Omega(u,v)
=
t(u-v) - e^{2ip_{\textbf{u}}(v)}t(v-u) \\ 
t(X) &= \frac{\mathrm{sh}(i\eta)}{\mathrm{sh}(X)\mathrm{sh}(X+i\eta)} \;\;\;\;\;\;\;\;
e^{2ip_{\mathbf{u}}(X)} =
\frac{1}{\prod_{m=1}^{L}
\frac{\mathrm{sh}( X-\theta_{m}+i\eta)}{\mathrm{sh}(X-\theta_{m})} }
\prod_{k=1}^{M}
\frac{\mathrm{sh}(X-u_{k}+i\eta)}{\mathrm{sh}(X-u_{k}-i\eta)} 
\end{align}
These results are simple generalizations of the XXX case obtained in \cite{Kostov1}; the quantity $\Omega$ is the so called Slavnov kernel \cite{Slavnov}, and $p_{\textbf{u}} $ is called the pseudo-momentum. It is  related to the Bethe equations by :
\begin{equation}
e^{2ip_{\mathbf{u}}(u)}
=
-1   \qquad  \forall u\in \mathbf{u} \;\;\;\;\; \text{satisfying Bethe equations}
\end{equation}
In order to make contact with the  more familiar  XXX case, we define the exponential variables :

\begin{equation}
q =e^{-2i\eta} \;\;\; z= e^{2u} \;\;\;  w = e^{2v} \;\;\;  \xi = e^{2\theta}
\end{equation}
and some condensed notations for the different quantities that will appear frequently in the following :

\begin{align}
 Q_{\mathbf{a}}(X) = \prod_{a\in \mathbf{a}}(X-a) \;\;\;\;
 Q_{\mathbf{a}}^{\pm}(X) = \prod_{a\in \mathbf{a}}(X-q^{\pm1}a)  \;\;\;\;
 E_{\mathbf{a}}^{\pm}(X) = \frac{Q_{\mathbf{a}}^{\pm}(X)}{Q_{\mathbf{a}}(X)}
\end{align}
Then it is possible  to express the Bethe equations and  the pseudo momentum in terms of $E_{\mathbf{z}}^{\pm}$. 
We write the function $t$ in terms of exponential variables and  factorize the Slavnov kernel in the following manner (the action of the operator $q^{z\partial_{z}}$ on a function results in the multiplication of its argument by $q$) :
\begin{equation}
\Omega(z,w)
=
(1-e^{2ip_{\textbf{z}}(w)}q^{-w\partial_{w}})
(q^{z\partial_{z}}-1)
\frac{2z}{z-qw}
\end{equation}

We can use the factorization of the Slavnov kernel to expand the factor $S_{\mathbf{u}\mathbf{v}}$ in  equation \eqref{innerprodxxz} :

\begin{equation}
\label{Sfactor}
S_{\mathbf{z},\mathbf{w}}=K
\frac{1}{\mathrm{det}_{jk}(\frac{1}{z_{j}-qw_{k}})}
\prod_{w\in \textbf{w}}(1-e^{2ip_{\textbf{z}}(w)}q^{-w\partial_{w}})
\prod_{z\in \textbf{z}}(q\,q^{z\partial_{z}}-1)
\mathrm{det}_{jk}(\frac{1}{z_{j}-qw_{k}})
\end{equation}
where
\begin{equation}
K
=
q^{M/2}\prod_{j=1}^{M}\sqrt{\frac{z_{j}}{w_{j}}}
\end{equation}

It appears that $S_{\mathbf{z},\mathbf{w}}$ is mainly the action of some differential operator on the Cauchy determinant $\mathrm{det}_{jk}(\frac{1}{z_{j}-qw_{k}})$. It is know that a Cauchy determinant can be splitted as a product of VanderMonde determinants, divided by  some polynomial. Using this fact here, and setting $V(\mathbf{a}) = \prod_{1\leq i < j \leq N}(a_{j}-a_{i}) $ we  get

\begin{align}
\label{SB}
S_{\mathbf{z},\mathbf{w}}
&=
(-1)^{M}K
\frac{1}{V(\mathbf{w})}
\prod_{w\in \textbf{w}}(1-q^{\frac{L}{2}}\frac{E_{\mathbf{z}}^{+}(w)}
{E_{\boldsymbol{\xi}}^{+}(w)}
q^{-w\partial_{w}}) V(\mathbf{w}) \times 
 \frac{1}{V(\mathbf{z})} \prod_{z\in \textbf{z}}(1-\frac{E_{\mathbf{w}}^{+}(z)}{q^{M-1}}q^{z\partial_{z}})
 V(\mathbf{z})
\end{align}

Looking carefully at the previous expression of $S_{\mathbf{z},\mathbf{w}} $, we see that a single relevant quantity appears twice. This motivates the difinition of the $B$ functional 
\begin{equation} 
B_{\mathbf{z}}^{\pm}[f]
=
\frac{1}{V(\mathbf{z})}
\prod_{z\in\mathbf{z}}\left(1-f(z)q^{\pm z\partial_{z}}\right)
V(\mathbf{z})
\end{equation}
Clearly, the $B$ functional is the XXZ analogue of the $A$ functional defined in \cite{Kostov1}.
We have to find functional equations satisfied by the $B$ functional to combine the two $B$ factors in the expression \eqref{SB} in a single $B$ functional defined on the set $\mathbf{z}\cup\mathbf{w} $.
We introduce the following  generic notation for partitions : $\mathbf{z}=\mathbf{z'}\cup\mathbf{z''}$, and perform an expansion on partitions :
\begin{equation}
B_{\mathbf{z}}^{\pm}[f]
=
\sum_{\mathbf{z'}\cup\mathbf{z''}}
\prod_{z''\in\mathbf{z''}}(-f(z'')q^{\pm(M-\frac{1}{2})})
\prod_{z''\in\mathbf{z''}}E_{\mathbf{z'}}^{\mp}(z'')  \;\;
q^{\mp\frac{1}{2}|\mathbf{z''}|^{2}}
\end{equation}
It is now possible to obtain the two functional equations on $B$ :
\begin{equation}
\label{funcB1}
B_{\mathbf{z}}^{+}[f]
=
B_{\mathbf{z}}^{-}[q^{2(M-1)} \; \left(\frac{1-q}{1-q^{-1}}\right)\;\frac{E_{\mathbf{z}}^{-}}{E_{\mathbf{z}}^{+}} f]
\;\;\;\;\;\;\;
B_{\mathbf{z}}^{-}[f]
=
B_{\mathbf{z}}^{+}[q^{-2(M-1)} \; \left(\frac{1-q}{1-q^{-1}}\right)^{-1}\;\frac{E_{\mathbf{z}}^{+}}{E_{\mathbf{z}}^{-}} f]
\end{equation}
Using the $B$ functional and the associated differential operator $\hat{B} $, the quantity $ S_{\mathbf{z},\mathbf{w}}$ is given by :
\begin{equation}
S_{\mathbf{z},\mathbf{w}}
=
(-1)^{M}K\;\;
\hat{B}_{\mathbf{w}}^{-}[q^{L/2}\frac{E_{\mathbf{z}}^{+}}{E_{\boldsymbol{\xi}}^{+}}]\cdot
B_{\mathbf{z}}^{+}[\frac{1}{q^{M-1}}E_{\mathbf{w}}^{+}]
\end{equation}
To conclude, we  need  to use the functional identities derived above, in the previous expression of $ S_{\mathbf{z},\mathbf{w}}$, as well as Bethe equations for the set $ \mathbf{z}$. This gives   a more symmetric expression for $S_{\mathbf{z},\mathbf{w}}$ :
\begin{equation}
S_{\mathbf{z},\mathbf{w}}
=
(-1)^{M}K\;\;
\hat{B}_{\mathbf{w}}^{-}[q^{L/2}\frac{E_{\mathbf{z}}^{+}}{E_{\boldsymbol{\xi}}^{+}}]\cdot
B_{\mathbf{z}}^{-}[q^{L/2}\frac{E_{\mathbf{w}}^{+}}{E_{\boldsymbol{\xi}}^{+}}]
\end{equation}
The symmetry of the previous formula allows  to obtain $S_{\mathbf{z},\mathbf{w}}$ as a single $B$ functional on the set $\mathbf{z}\cup\mathbf{w}$ :
\begin{equation}
S_{\mathbf{z},\mathbf{w}}
=
(-1)^{M}K\;\; B_{\mathbf{z}\cup\mathbf{w}}^{-}\left[\frac{q^{\frac{L}{2}+M}}{E^{+}_{\boldsymbol{\xi}}}
\right]
\end{equation}
The final result for the scalar product is then :
\begin{equation}
\label{final2}
\langle \textbf{v}|\textbf{u}\rangle
=
\prod_{j=1}^{M}
\mathcal{A}(v_{j})\mathcal{D}(u_{j})
(-1)^{M}K\;\; B_{\mathbf{z}\cup\mathbf{w}}^{-}\left[\frac{q^{\frac{L}{2}+M}}{E^{+}_{\boldsymbol{\xi}}}
\right]
\end{equation}
where both usual and exponential variables have been used. The result is  similar to the expression for the XXX case, derived in \cite{Kostov1}. The next section outline the investigation of the $B$ functional in the semiclassical limit, done in \cite{Kostov2} for the XXX case.

\section{Semi-classical expansion of the scalar products}
We aim now to study the situation where the number of rapidities is taken to be infinitely large.
We restrict ourselves to the study of the quantity $ B_{\mathbf{y}}^{-}[f]$, where $\mathbf{y} $ is a set of complex numbers that do not necessarily satisfy the Bethe equations, and $f$ is a rational  function depending on the inhomogeneities. We can also suppose that the elements of $\mathbf{y} $ are located on a macroscopical arc in the complex plane, and their number is finite. Following the steps in \cite{Kostov2}, we aim to express $B_{\mathbf{y}}^{-}[f] $ as an integral. The representation of the  Izergin-Korepin determinant (which is an analytical result for the Bethe scalar product \eqref{innerprod}) by shift operators (of type $q^{z\partial_{z}}$) was done in \cite{Warnaar}. Here we obtain :
\begin{equation}
B_{\mathbf{y}}^{-}[f]
=
\sum_{\alpha}(-1)^{\alpha}\det
\left( 
\frac{E_{j}y_{j}(1-q^{-1})}{y_{j}-q^{-1}y_{k}}
\right)
\end{equation}
where we have set : $E_{j}=f(y_{j})\prod_{k\neq j} \frac{y_{k}-q^{-1}y_{j}}{y_{k}-y_{j}}$. This can be combined to :
\begin{equation}
\label{Bdet}
B_{\mathbf{y}}^{-}[f]
=
\det(I-K)
\end{equation}
with $K$ the $N\times N$ matrix :
\begin{equation}
K_{jk}
=
\frac{E_{j}y_{j}(1-q^{-1})}{y_{j}-q^{-1}y_{k}}
\end{equation}
Then representing the determinant with fermions \cite{Kostov2} (normalised by $\langle\psi(x)\psi^{\star}(y)\rangle=\frac{1}{x-y}$) :
\begin{equation}
B_{\mathbf{y}}^{-}[f]
=
\langle 0 |
\exp \left( \sum_{i=1}^{N} E_{j}y_{j}(1-q^{-1})\psi^{\star}(y_{j})\psi(q^{-1}y_{j})
 \right)
| 0 \rangle
\end{equation}
Which can be rewritten as :
\begin{equation}
B_{\mathbf{y}}^{-}[f]
=
\langle 0 |
\exp \left( - \int_{C_{\mathbf{y}}} \frac{\mathrm{d}x}{2\pi i} H_{q}(x) \psi^{\star}(x)\psi(q^{-1}x) \right)
| 0 \rangle
\end{equation}
where the contour $C_{\mathbf{y}}$ encircles  the complex numbers $\mathbf{y}$ and 
\begin{equation}
H_{q}(X)
:=
f(X)\prod_{i=1}^{N}\frac{q^{-1}X-y_{k}}{X-y_{k}}
\end{equation}
Representing fermion fields as vertex operators ($\psi(x)=e^{\phi(x)}\;\psi^{\ast}(x)=e^{-\phi(x)}$) and calculating their correlation functions, we get
\begin{equation}
\label{final3}
B_{\mathbf{y}}^{-}[f]
=
\sum_{n=0}^N
\frac{(-1)^n}{n!}
\prod_{j=1}^n
\oint_{C_{\mathbf{y}}}\frac{\mathrm{d}x_{j}}{2\pi i}
\left(\frac{H_{q}(x_{j})}{-(1-q^{-1})x_{j}}\right)
\prod_{k>j}^{n}
\frac{(x_{j}-x_{k})^{2}}{(x_{j}-qx_{k})(x_{j}-q^{-1}x_{k})}
\end{equation}
Indeed, the factors in the second product are  the correlation functions of two vertex operators $\mathcal{V}_{q}(x) $ :
\begin{equation}
\mathcal{V}_{q}(x)=e^{\phi(q^{-1}x)-\phi(x)}
\end{equation}
which are given by :
\begin{equation}
\langle 0 |
\mathcal{V}_{q}(x)
\mathcal{V}_{q}(y)
| 0 \rangle
=\frac{(x-y)^{2}}{(x-qy)(x-q^{-1}y)}
\end{equation}
This integral formula for $B_{\mathbf{y}}^{-}[f]$ is reminiscent of the grand canonical partition function for a gas of particles in a pairwise interaction potential. The previous expression of $B_{\mathbf{y}}^{-}[f]$ is therefore  well adapted to perform a cluster expansion \cite{Tong}.
Define :
\begin{equation}
1+f_{jk}
=
\frac{(x_{j}-x_{k})^{2}}{(x_{j}-qx_{k})(x_{j}-q^{-1}x_{k})}
\end{equation}
We set $Z_{n} $ to be the canonical partition function of the gas of particles :
\begin{equation}
Z_{n}
=
\frac{(-1)^{n}}{n!}
\prod_{j=1}^n
\oint_{C_{\mathbf{y}}}\frac{\mathrm{d}x_{j}}{2\pi i}
\left(\frac{H_{q}(x_{j})}{-(1-q^{-1})x_{j}}\right)
\prod_{j<k}
(1+f_{jk})
\end{equation}
The idea of the cluster expansion is the following : we expand the product  $\prod_{j<k}(1+f_{jk}) $ as a sum, whose terms are represented by graphs. To each graph we associate a weight (the value of the integral of the corresponding term). This gives : 
\begin{equation}
Z_{n}
=
\frac{(-1)^{n}}{n!}
\sum_{G} W[G]
\end{equation}
where the summatio is performed on all possible graphs $G$ with weights $W[G]$. We now apply the cluster expansion of \cite{Tong}. Defining $U_{l} $ to be :
\begin{equation}
U_{l}
=
\int_{C_{\mathbf{y}}}\prod_{i=1}^{l}\frac{\mathrm{d}x}{2\pi i} 
\left( \frac{H_{q}(x_{j})}{-(1-q^{-1})x_{j}} \right)
\sum_{G \in {\text{ $l$-cluster} }} W[G]
\end{equation}
The cluster expansion yields :
\begin{equation}
\label{final3}
Z_{n}
=
(-1)^{n}
\sum_{\lbrace m_{l} \rbrace }
\prod_{l}\frac{U_{l}^{m_{l}}}{(l!)^{m_{l}}m_{l}!}
\end{equation}
where the sum is calculated with the constraint $\sum_{l=1}^{n}m_{l}\, l = n $.
The final result for $B_{\mathbf{y}}^{-}[f]$ is therefore :
\begin{equation}
\label{final}
B_{\mathbf{y}}^{-}[f]
=
\sum_{n=0}^{N}Z_{n}
\end{equation}
We would like to compute this sum in the limit $N\rightarrow\infty $, ie when we can relax the constraint $\sum_{l=1}^{n}m_{l}\, l = n  $. In this case, $\sum_{n=0}^{N}Z_{n} $ can be written in an elegant way as an exponential of a certain power series \cite{Tong}. This is what we would like to do, but we have to cope with a major obstruction : the term $U_{l} $ depends on the number $N$ through the function $H_{q} $. Hence, it appears to be difficult to find the limit $N\rightarrow\infty $ of $B_{\mathbf{y}}^{-}[f]$ from the relation \eqref{final}, \textit{whithout assuming precise constraints on the set of rapidities} \textbf{y}.
Nevertheless it is possible to formulate a conjecture on what the leading term should be : since in the XXX case the leading term of the scalar product was essentially  given by the exponential of the dilogarithm function \cite{Kostov2}, we can expect for the XXZ case the same kind of identity by involving this time the quantum dilogarithm  :

\begin{equation}
\log
\langle \textbf{v}|\textbf{u}\rangle
\sim
\int_{C_{\mathbf{u}\cup \mathbf{v}}}
\frac{\mathrm{d}x}{2\pi}
\log\Psi(e^{ip_{u}(x)+ip_{v}(x)})
\end{equation}
where the contour $C_{\mathbf{u}\cup \mathbf{v}}$ encircles the rapidites $\mathbf{u}$ and $\mathbf{v}$ and quantum dilogarithm is defined by :
\begin{equation}
\Psi(x)
=
\prod_{n=0}^{\infty}(1-xq^{n}) \quad |q|<1
\end{equation}
The motivation for this conjecture is the following : it is possible to recover the XXX spin chain from the XXZ spin chain by sending the parameter $\eta$ to $0$. On the other hand, if we set $q =e^{\eta}$ and set $\eta\rightarrow0 $ we have :
\begin{equation}
\Psi(x)
=
\frac{1}{\sqrt{1-x}}
e^{\frac{1}{\eta}\mathrm{Li}_{2}(x)} (1+ \mathcal{O}(\eta))
\end{equation}
which involves the same expression that was found for the leading term in the semiclassical expansion of the XXX case in \cite{Kostov2}.

\section{Conclusion}
In this note, our main result \eqref{final2} allows to write a XXZ Bethe scalar product in terms of a specific quantity: the $B$ functional, which is well designed for the semiclassical expansion. The natural way to continue our investigation of the XXZ inner product is to understand the structure of the Bethe roots in this case. This could give a hint on which exact hypothesis we should fix on the set of rapidities $\mathbf{y} $ in order to obtain a manageable mathematical calculation for the semiclassical expansion. 
\\
\\
\textbf{
Acknowledgements 
}
\\
I would like to thank Ivan Kostov and Didina Serban, who supervised me during my internship at IPhT, proposed this project and guided me through my work. I am also grateful to Andrei Petrovskii for valuable discussions.
\\
\\
\textbf{Note added}
\\
After this research project was completed, the paper \cite{Jiang} by Y. Jiang and J. Brunekreef appeared where the same problem has been adressed.

\end{small}

\end{document}